# Realization of chiral whispering gallery mode cavities enabled by photonic Chern insulators

Hao-Chang Mo, Zi-Xuan Gao, Xiao-Dong Chen[*], Jian-Wen Dong

*School of Physics & State Key Laboratory of Optoelectronic Materials and Technologies,
Sun Yat-sen University, Guangzhou 510275, China.*

[*]Corresponding author: chenxd67@mail.sysu.edu.cn

## ABSTRACT

Recently, whispering gallery modes (WGMs) have attracted considerable attention due to their extensive applications in the development of on-chip microcavities, high-sensitivity sensors, and high-performance lasers. Conventional WGMs are achiral under the time-reversal symmetry, and show high sensitivity to defects in optical devices. Here, we introduce topological physics into photonic cavities and demonstrate the realization of chiral WGMs enabled by photonic Chern insulators. Through comprehensive numerical simulations and experimental measurements, we reveal the critical differences between chiral and achiral WGMs, highlighting the robustness of chiral WGMs even in the presence of defects within the cavities. Our research provides valuable insights into the design of robust WGM cavities and offers a novel platform for exploring light-matter interaction phenomena.

*Introduction*. --- Over the past century, whispering gallery modes (WGMs) have been extensively explored and investigated in acoustics and optics.[1-15] In a closed structure (e.g., the Whispering Gallery in St. Paul's Cathedral in London and the Hall of Prayer for Good Harvests in the Temple of Heaven in Beijing), acoustic waves could reflect continuously off the smooth walls, allowing them to propagate over long distances with minimal loss. In optics, WGMs originate from the total internal reflection at the interface between medias with different refractive index.[3,16] In a photonic cavity which supports WGMs, light beams are localized in the cavity, maintaining stable traveling wave modes. Due to high quality factors and small mode volumes, optical WGMs have been widely used in optical sensing, quantum optics and nonlinear optics, showing significant potential for integrated optical devices.[10,12] Under time-reversal symmetry, traditional optical WGM cavities are achiral, i.e., supporting both clockwise and anticlockwise WGMs. When fabrication defects and structural irregularities are introduced, WGMs are sensitive, often leading to observable changes in the transmission spectra with mode shift, mode splitting or mode broadening.[13-15]

Recently, the emergence of topological photonics has revolutionized the design of high-performance photonic devices such as waveguides, cavities, splitters and couplers.[17-30] By designing photonic systems, such as photonic crystals (PCs), to exhibit nontrivial topology, researchers have enabled the realization of topologically protected photonic states which exhibit robustness to fabrication defects and structural irregularities. Among various topological photonic states, the topological edge states are extensively studied as they are immune to backscattering and perturbations, and have found applications in one-way waveguide, nonreciprocal filter and robust topological laser.[31-41] In the context of enhancing optical signal transmission, incorporating topological edge states into the design of cavities that support topologically protected whispering gallery modes (WGMs) presents a promising avenue. Previous research has successfully demonstrated WGMs with different pseudospins using the optical quantum spin Hall effect.[20,42,43] These cavities, which preserve time-reversal symmetry and exhibit low transmission loss, have been applied in areas such as quantum electrodynamics and high-performance lasers. Moreover, by employing gyromagnetic materials to break time-reversal symmetry, researchers can realize chiral edge states (CESs), a specific type of topological edge state.[31,32,34-36] This approach has led to the theoretical proposal of chiral WGMs based on CESs.[44] These WGMs are characterized by non-degenerate modes with unidirectional propagation, which support traveling wave modes exclusively, without generating standing wave modes, and exhibit high insensitivity to defects. Despite the promising theoretical groundwork, the experimental observation of such WGMs remains elusive, casting uncertainty on their practical applications and highlighting an area of ongoing research.

In this work, we report the experimental realization of a chiral WGM cavity based on a photonic Chern insulator. Using yttrium iron garnet (YIG) ferrite cylinders with and without external magnetic fields, we design two different types of two-dimensional photonic crystals (2D PCs), i.e., the photonic Chern insulator and the photonic trivial insulator, and subsequently construct the chiral and achiral WGM cavities. To characterize the chiral WGM cavity, we perform the eigenfrequencies simulation, and experimentally conduct the pump-probe measurement. As a comparison, an achiral WGM cavity is constructed with the photonic trivial insulator and experimentally characterized in a similar manner. After introducing similar defects into both cavities, we provide the convincing evidence for the robust nature of the chiral WGMs by measuring the transmission spectra. Our research offers a feasible technique to realize a chiral WGM cavity, which could be a robust platform for the development of high-efficiency filters and topological photonic circuits.

*Chiral and achiral WGM cavities*. --- As illustrated in Fig. 1(a), the chiral WGM cavity is constructed by surrounding a photonic Chern insulator (characterized by a nonzero Chern number of $C = 1$) with a photonic ordinary insulator (characterized by $C = 0$). According to the bulk-edge correspondence[17,18], gapless CES can exist at the interface between these two photonic insulators with different Chern numbers. Consequently, in a closed cavity formed by encircling this interface, topologically protected chiral WGMs will be present [right panel of Fig. 1(a)]. The frequencies of chiral WGMs are determined by the effective length of the cavity and the CES dispersion curve, and the direction of energy flow of chiral WGMs is consistent with that of the

CES. In contrast, an achiral WGM cavity is constructed by photonic ordinary insulators [Fig. 1(b)]. The cavity can be formed by enclosing a traditional photonic waveguide between two photonic ordinary insulators. Since traditional photonic waveguides support trivial edge states, the achiral cavity supports frequency-degenerate clockwise and anticlockwise WGMs [right panel of Fig. 1(b)]. If defects are introduced into the chiral WGM cavity, the frequency of the topologically protected chiral WGMs will shift but not split into two modes, as illustrated in Fig. 1(c). In contrast, the existence of both clockwise and anticlockwise WGMs making the achiral cavity more sensitive to defects. The same types of defects can lead to mode frequency shifts and mode splitting, as illustrated in Fig. 1(d). To validate the theoretical predictions, we will start with the design of photonic Chern insulator and photonic ordinary insulator, followed by the experimental design and measurement of chiral and achiral WGM cavities.

*Design of chiral and achiral WGM cavities.* --- Initially, we design a photonic Chern insulator and a photonic ordinary insulator using 2D PCs with transverse-magnetic modes. The 2D gyromagnetic photonic crystal (GPC) is constructed with the commercially available YIG ferrites. Under a nonzero out-of-plane magnetic field $B_0$, the in-plane relative permeability tensor of YIG ferrites has the form of $\boldsymbol{\mu} = \begin{pmatrix} \mu_r & i\mu_\kappa \\ -i\mu_\kappa & \mu_r \end{pmatrix}$ where $\mu_r = 1 + \frac{(\omega_0 - i\alpha\omega)\omega_m}{(\omega_0 + i\alpha\omega)^2 - \omega^2}$, $\mu_\kappa = \frac{\omega\omega_m}{(\omega_0 + i\alpha\omega)^2 - \omega^2}$, $\omega_m = \gamma\mu_0 M_s$, $\omega_0 = \gamma\mu_0 H_0$, $H_0 = \frac{B_0}{\mu_0}$. Here, $\gamma = 1.759 \times 10^{11}$ C/kg is the gyromagnetic ratio, $\alpha = 0.0002$ is the damping coefficient, and the saturation magnetization $M_s$ satisfies $4\pi M_s = 1950$ Gs. The relative permittivity of YIG ferrites at this stage is $\varepsilon_r = 14.6$. The GPC consists of a honeycomb lattice of YIG rods (diameter $d = 6$ mm) in the air, with the unit cell depicted in the lower inset of Fig. 2(a). The bulk bands of this GPC under an external magnetic field of $B_0 = 0.043$T are shown in Fig. 2(a). The effective Hamiltonian for the designed photonic crystal has the form of $H_{eff} = v_D(\delta k_x \sigma_x + \delta k_y \sigma_y) + m_T \sigma_z$, where $\delta \boldsymbol{k} = (\delta k_x, \delta k_y)$ measures from the K point, $\sigma_i$ ($i = x, y, z$) are the Pauli matrices, $v_D$ is the group velocity, $m_T$ is the effective mass induced by the breaking of time-reversal symmetry.[45,46] This effective Hamiltonian is fundamentally different from that of inversion-symmetry-breaking honeycomb photonic crystals that preserve time-reversal symmetry. As the time-reversal symmetry is broken, a bandgap spanning from 5.02 to 5.49 GHz is found. The gap Chern number, calculated by summing the Chern numbers of all bands below the gap[24,34,47], is $C = 1$, indicating that this GPC functions as a photonic Chern insulator. The Chern number of a given band can be numerically obtained by two different methods. The first method involves integrating the Berry curvature over the entire Brillouin zone, i.e., $C = \frac{1}{2\pi} \int_{BZ} i\nabla \times \langle E_z(\mathbf{k}) | \nabla_\mathbf{k} | E_z(\mathbf{k}) \rangle d^2k$ where $E_z(\mathbf{k})$ is the eigen wave function at the momentum $\mathbf{k}$ [24]. The other method evaluates the eigenvalues of symmetry operators at high-symmetry points, i.e., $\exp(i\pi C/3) = \prod \eta(\Gamma)\theta(K)\zeta(M)$ where $\eta(\Gamma)$, $\theta(K)$ and $\zeta(M)$ represent the eigenvalues of the rotation operators $\hat{C}_6$, $\hat{C}_3$ and $\hat{C}_2$ on bulk modes at the $\Gamma$, K and M points, respectively [47]. Meanwhile, we design a 2D triangular PC consisting of the identical YIG rods in the air. Without an external magnetic field, the relative permittivity of the YIG ferrites is $\varepsilon_r = 10.2$.[48] The corresponding bulk band has a bandgap between 5.19 and 7.91 GHz, characterized by a gap Chern number $C = 0$ [Fig. 2(b)], classifying this PC as a photonic ordinary insulator. It is noteworthy that both types of PCs share an identical lattice constant of $a$ = 20mm and are composed of the same YIG rods, facilitating the design and realization of the chiral and achiral WGM cavities.

Next, we numerically study the edges constructed by both photonic insulators. The bulk-edge correspondence indicates that an interface between the photonic Chern insulator and the photonic ordinary insulator will support CESs, with the number and group velocity of CESs depending on the difference of the gap Chern numbers across the interface. We consider an edge between the nontrivial GPC and the trivial PC [right inset of Fig. 2(c)]. The calculated band dispersion depicted in Fig. 2(c) shows the presence of one gapless CES with positive group velocity in the bandgap. This observation consists with the difference of the gap Chern numbers across the interface, $\Delta C = 1$. Conversely, the W1 photonic waveguide, formed by removing a row of rods from the otherwise-perfect trivial PC [right inset of Fig. 2(d)], does not support CESs. The projected band shows a pair of edge states with opposite group velocities [Fig. 2(d)].

Finally, we design and construct the WGM cavities based on the above edges. As illustrated in Fig. 2(e),

the chiral WGM cavity features a regular hexagonal structure with a side length of $L = 24a$. We numerically calculate the frequencies of cavity modes and plot them in Fig. 2(g). Consistent with the projected band structure depicted in Fig. 2(b), only chiral WGMs are found in the bandgap. The spatial profile $|E_z|^2$ of one representative chiral WGM [labeled by a circled number ①] clearly demonstrates that the energy is localized inside the cavity. The time-averaged Poynting vectors [indicated by cyan arrow-cones] indicate that the energy flows in a clockwise direction, confirming the presence of a chiral WGM. Similarly, we construct a hexagonal achiral WGM cavity [Fig. 2(f)] with the same side length as the chiral WGM cavity and perform the eigenfrequencies calculation. As depicted in Fig. 2(h), doubly degenerate achiral WGMs are found within the bandgap due to the hexagonal symmetry of the cavity. The spatial profiles $|E_z|^2$ of a pair of these clockwise and anticlockwise WGMs [marked by ② and ③] are shown in the subfigures of Fig. 2(i). Although the energy distributions of the degenerate achiral WGMs are analogous, their energy flows are in opposite directions. This observation is consistent with the fact that the W1 waveguide simultaneously supports two edge states with opposite group velocities. Additionally, the subfigure labeled as ④ in Fig. 2(i) illustrates one of the non-degenerate modes. While its energy is also localized in the cavity, the Poynting vectors reveal that this mode is not a traveling wave mode and, therefore, is not considered further in this paper. Notably, ordinary cladding layers are typically used to prevent energy leakage from the cavity. However, if the chiral edge modes lie below the light cone [34,36], such cladding layers can be omitted, representing a significant advancement for device integration by minimizing the required space.

*Realization of chiral and achiral WGM cavities.* --- The schematic of the fabricated experimental sample is shown in Fig. 3(a).[49,50] Initially, we use two aluminum plates with a thickness of $h_1 = 1$mm to construct a planar waveguide with a height of $h = 5$ mm, which supports transverse magnetic modes in the microwave frequency range of interest. As depicted in the upper right panel of Fig. 3(a), we construct the GPC by arranging the YIG rods with a height of $h$ and a diameter of $d_1 = 6$ mm in a honeycomb lattice inside the planar waveguide. To induce an external magnetic field, we sandwich the planar waveguide between other two aluminum plates which are embedded with permanent magnet rods. These magnets, each with a thickness of $h_2 = 2$ mm and a diameter of $d_1$, are positioned directly above and below the YIG rods, thereby creating an effective magnetic field along the YIG rods. The color of magnets indicates that the magnetic field is applied along the positive direction of the z-axis. The construction of the trivial PC involves arranging the YIG rods in a triangular lattice inside the planar waveguide, as shown in the lower right panel of Fig. 3(a). In contrast to the configuration of the GPC, this setup does not require any magnets. To insert the source and probe antenna for measurement, all aluminum plates are drilled with sampling holes with a diameter of $d_2 = 2$mm. These sampling holes are arranged in a rectangular lattice with lattice constants $a_x = a/2$ and $a_y = \sqrt{3}a/4$. Additionally, their positions are meticulously chosen to avoid any collision with the YIG rods and magnets. In comparison to the lattice constant $a$, the diameter of the sampling holes is sufficiently small, ensuring that they have no discernible impact on our conclusions.

The fabricated chiral WGM and achiral WGM cavities are depicted in Figs. 3(b) and 3(c), respectively. To experimentally characterize the WGMs, we conduct the pump-probe measurements on these cavities using a vector network analyzer (Anritsu ShockLine™ MS46122B) and a pair of dipole antennas (serve as the source antenna and probe antenna, respectively). A detailed demonstration of the measurement setup is shown in Fig. 3(f). The positions of the source and probe antennas are marked by the filled and unfilled stars, which are consistent with those presented in Figs. 3(g) and 3(h), respectively. The measured $|E_z|$ spectrum of the chiral WGM cavity, as depicted in Fig. 3(d), has four peaks within the bandgap, which is consistent with the simulation result in Fig. 2(g). Similarly, the pump-probe measurement result for the achiral WGM cavity, as depicted in Fig. 3(e), also matches well with the simulation results depicted in Fig. 2(h).

To elucidate the distinct difference between chiral and achiral WGM cavities, we introduce defects into both cases and perform pump-probe measurements to observe the response of the WGMs. The introduced defects will break the symmetry of the cavity and change its effective length, and thus affect mode properties. For a clearer comparison, only one chiral WGM or a pair of achiral WGMs are considered in the subsequent analysis. As depicted in the upper panel of Fig. 4(a), the first defect (labeled as defect 1) is introduced by removing two YIG rods in the GPC near the cavity. The resonant peak shifts to the lower frequency, indicating that the effective length of the cavity is increased by this defect [lower panel of Fig. 4(a)]. The second defect [labeled as defect 2] is introduced by removing the other two YIG rods in the GPC, which differs from the defect 1 in missing rod positions [upper panel of Fig. 4(b)]. Despite the violation of the cavity's symmetry by the defect 2, the chiral WGM still unidirectionally propagates around the defect. Therefore, it also only changes the effective length of the cavity and shifts the resonant frequency of the chiral WGM. We also combine the defect 1 and defect 2 to introduce the last defect [labeled as defect 1 & 2 in Fig. 4(c)]. The pump-probe measurement result, depicted in the lower panel of Fig. 4(c), indicates that the WGM frequency has been shifted to lower value. Due to the competing effects of defect 1 and defect 2 on the resonant frequency of the WGM, the resulting frequency shift is smaller than what would be observed if only defect 1 were present. Despite introducing three different defects into the chiral WGM cavity, the nontrivial topology of the photonic Chern insulator ensures that these defects only cause frequency shifts in the chiral cavity. To see this, we summary the resonant frequency of the chiral WGM of four different cavities [Fig. 4(d)]. Under the introduced defects, the chiral WGM preserves the single resonant peak, since its presence does not depend on the cavity's symmetry which is broken by defects. Additionally, we calculate the free spectra range (FSR) of the studied chiral WGM WGM.[16,51] As shown in Fig. 4(e), the FSR of the chiral WGM remains stable under different defects.

Similarly, by removing YIG rods of the trivial PC near the cavity, we introduce the aforementioned three types of defects into the achiral WGM cavity. The corresponding results of pump-probe measurements are depicted in Figs. 4(f)-4(h). Comparing to the results in the chiral WGM cavity, we observe that introducing defects in the achiral WGM cavity not only causes mode shift but also leads to mode splitting. This originates from the fact that a pair of the doubly degenerate WGMs have different responses to the defects. In the defect-free achiral WGM cavity, the clockwise and anticlockwise WGMs are degenerate under the protection of rotation symmetry of the cavity. The field distributions of degenerate modes are also symmetric, making it difficult for them to couple. When the symmetry is broken, the fields of the degenerate modes are perturbed and no longer strictly symmetric. As a result, the coupling between modes is enhanced, leading to mode splitting. Therefore, the degenerate WGM splits into two non-degenerate modes with different frequencies, which manifests as a single resonance peak splitting into two resonance peaks [lower panels of Figs. 4(f) and 4(g)]. Meanwhile, the split modes induced by defects also experience frequency shifts due to the mode coupling and the change in cavity's length. The mode coupling effect induced by defects leads to mode splitting, which reduces the coupling efficiency and quality factor, thereby increasing the design complexity of achiral cavities. When the combination of the defect 1 and defect 2 is introduced into the cavity, the degenerate achiral WGMs will split into two modes as well. Additionally, the different responses of the clockwise and anticlockwise WGMs to defects results in a significantly small peak compared to the other, as shown in Fig. 4(h). For a clear demonstration, we summary the performance of achiral WGM cavities to different defects. As shown in Fig. 4(i), the achiral WGM exhibits mode splitting under the introduced defects. We also calculate the FSR of the studied achiral WGM [Fig. 4(j)]. The achiral WGM exhibits a larger FSR in the perfect cavity due to the wider bandgap. However, once defects are introduced into the cavity, the achiral WGM suffers from mode splitting, leading to a sharp decrease of FSR. The measured FSR of the chiral and achiral WGM are shown in the Table 1 below. It is consistent with that presented in Figs. 4(e) and 4(j). For the chiral WGM, the FSR remains stable (varying from 52 to 66 or 55). In contrast, for the achiral WGM, the FSR decreases by an order of magnitude (drop from 271 to 38 or 48).

Table 1 | Free spectral range (FSR) of chiral and achiral WGM.

|  | Perfect | Defect 1 | Defect 2 | Defect 1 & 2 |
|---|---|---|---|---|
| Chiral WGM | 52 | 66 | 55 | 55 |
| Achiral WGM | 271 | 38 | 48 | 48 |

Note: FSR values are given in MHz.

Apart from different defects in the cavity, we also investigate the influence of irregularities, which are commonly seen in the fabrication of resonant cavities, on the WGMs. We consider three different shapes which are labeled as shape 1-3. These shapes are constructed by inwardly recessing the trivial PC surrounding the cavity inward while maintaining an approximately constant effective cavity length of $L = 24a$. The depth of recess represents the gradually increasing irregularity in the cavity. As shown in Fig. 5, we experimentally construct chiral [Figs. 5(a)–5(c)] and achiral WGM cavities [Figs. 5(f)–5(h)] with the gradually increasing irregularity. For a better comparison, we place the dipole antennas at the same positions as depicted in Fig.3 on both the chiral and achiral WGM cavities with introduced irregularities to perform the pump-probe measurements. The corresponding measured $|E_z|$ spectra are plotted in the lower panel of Figs. 5(a)–5(c) and Figs. 5(f)-5(h), respectively. The measured results indicate that the introduced irregularities do not disrupt the presence of the chiral WGM but do cause a frequency shift. In contrast, measured $|E_z|$ spectra for the achiral WGM cavity show that the achiral WGM exhibits both mode splitting and frequency shift. These conclusions are further supported by the resonant frequency and calculated FSR of chiral and achiral WGM depicted in Figs. 5(d)-5(e) and Figs. 5(i)-5(j), respectively. In comparison, the protection provided by nontrivial topology endows chiral WGMs with significantly enhanced robustness compared to conventional WGMs.

*Conclusion and outlook.* --- We experimentally realize a chiral WGM cavity by incorporating two types of PCs, i.e., the photonic Chern insulator and the photonic ordinary insulator. By applying CES of the photonic Chern insulator, we obtain chiral WGMs. Subsequently, we characterize the chiral WGMs using the pump-probe measurement. For comparison, we construct and experimentally characterize an achiral WGM cavity based on the photonic ordinary insulator. By introducing similar defects in both cavities and measuring the response spectra of the cavities with these defects, we confirm the robustness of chiral WGMs. As discussed above, when the defects are too small to form local cavity modes, the chiral modes will stably exist. If the defects are sufficiently large (e.g., removing half of the unit-cells near the interface between PCI and POI), they will form local trivial cavities, leading to localization of chiral modes. Such defects will violate the structure of cavity and reduce the quality factors of the chiral modes. However, since such "defects" are easier to be avoided in the fabrication of the cavity compared to the small ones, our design remains sufficiently practical. Our research provides a new approach for the realization and investigation of robust WGMs. Notably, beyond the hexagonal structure, our design methodology enables the chiral WGM cavity to be constructed in various closed shapes (e.g., rhombus or trapezium) to meet different requirements. Based on scaling laws of Maxwell equations, our design methodology can be conveniently adapted to terahertz and optical frequencies with appropriate material selection and design.[52,53] Although the magneto-optical effects of ordinary materials are weak in terahertz and optical frequencies, researchers have successfully realized strong magneto-optical effects in base on La: YIG and RIG film in terahertz region and van der Waals magnet and Fabry–Pérot cavity in optical region.[54-57] In the future, researches on extending the chiral WGM to the terahertz and optical frequency will introduce new candidates for defect-sensitive photonic devices, enabling the development of highly stable optical communication systems, quantum information processing systems, and other critical optical systems that demand high robustness.

**Acknowledgements**

This work was supported by National Key Research Development Program of China (2022YFA1404304), National Natural Science Foundation of China (12374364, 62035016), Guangdong Basic and Applied Basic Research Foundation (2023B1515040023), Guangzhou Science, Technology and Innovation Commission (2024A04J6333), Fundamental Research Funds for the Central Universities of the Sun Yat-sen University (23lgbj021).

# REFERENCES

[1] L. Rayleigh. CXII. The problem of the whispering gallery. *The London, Edinburgh, and Dublin Philosophical Magazine and Journal of Science*. **20**, 1001 (1910).
[2] R. D. Richtmyer. Dielectric Resonators. *J. Appl. Phys.* **10**, 391 (1939).
[3] M. Born and E. Wolf. *Principles of optics: electromagnetic theory of propagation, interference and diffraction of light* (Elsevier, 2013).
[4] C.-L. Zou, C.-H. Dong, J.-M. Cui, F.-W. Sun, Y. Yang, X.-W. Wu, Z.-F. Han, and G.-C. Guo. Whispering gallery mode optical microresonators: fundamentals and applications *SCIENTIA SINICA Physica, Mechanica & Astronomica* **42**, 1155 (2012).
[5] F. Vollmer and L. Yang. Review Label-free detection with high-Q microcavities: a review of biosensing

mechanisms for integrated devices. *Nanophotonics*. **1**, 267 (2012).

[6] E. Kim, M. D. Baaske, and F. Vollmer. Towards next-generation label-free biosensors: recent advances in whispering gallery mode sensors. *Lab on a Chip*. **17**, 1190 (2017).

[7] Y. Zhi, X.-C. Yu, Q. Gong, L. Yang, and Y.-F. Xiao. Single Nanoparticle Detection Using Optical Microcavities. *Advanced Materials*. **29**, 1604920 (2017).

[8] S.-J. Tang, B.-B. Li, and Y.-F. Xiao. Optical sensing with whispering-gallery microcavities. *PHYSICS*. **48**, 137 (2019).

[9] L. Cai, J. Pan, Y. Zhao, J. Wang, and S. Xiao. Whispering Gallery Mode Optical Microresonators: Structures and Sensing Applications. *Physica Status Solidi a-Applications and Materials Science*. **217**, 1900825 (2020).

[10] X. Jiang, A. J. Qavi, S. H. Huang, and L. Yang. Whispering-Gallery Sensors. *Matter*. **3**, 371 (2020).

[11] Y. Chen, Y. Yin, L. Ma, and O. G. Schmidt. Recent Progress on Optoplasmonic Whispering-Gallery-Mode Microcavities. *Advanced Optical Materials*. **9**, 2100143 (2021).

[12] D. V. Strekalov, C. Marquardt, A. B. Matsko, H. G. L. Schwefel, and G. Leuchs. Nonlinear and quantum optics with whispering gallery resonators. *Journal of Optics*. **18**, 123002 (2016).

[13] W. Liu, Y.-L. Chen, S.-J. Tang, F. Vollmer, and Y.-F. Xiao. Nonlinear Sensing with Whispering-Gallery Mode Microcavities: From Label-Free Detection to Spectral Fingerprinting. *Nano Lett*. **21**, 1566 (2021).

[14] N. Toropov, G. Cabello, M. P. Serrano, R. R. Gutha, M. Rafti, and F. Vollmer. Review of biosensing with whispering-gallery mode lasers. *Light Sci. Appl*. **10**, 42 (2021).

[15] J. Liu *et al.* Emerging material platforms for integrated microcavity photonics. *Science China Physics, Mechanics & Astronomy*. **65**, 104201 (2022).

[16] V. Van. *Optical microring resonators: theory, techniques, and applications* (CRC Press, 2016).

[17] L. Lu, J. D. Joannopoulos, and M. Soljačić. Topological photonics. *Nat. Photonics*. **8**, 821 (2014).

[18] T. Ozawa *et al.* Topological photonics. *Reviews of Modern Physics*. **91** (2019).

[19] A. B. Khanikaev and G. Shvets. Two-dimensional topological photonics. *Nat. Photonics*. **11**, 763 (2017).

[20] G. Siroki, P. A. Huidobro, and V. Giannini. Topological photonics: From crystals to particles. *Phys. Rev. B*. **96**, 041408 (2017).

[21] B. Y. Xie, H. F. Wang, X. Y. Zhu, M. H. Lu, Z. D. Wang, and Y. F. Chen. Photonics meets topology. *Opt Express*. **26**, 24531 (2018).

[22] N. A. Mortensen, S. I. Bozhevolnyi, and A. Alù. Topological nanophotonics. *Nanophotonics*. **8**, 1315 (2019).

[23] M. Kim, Z. Jacob, and J. Rho. Recent advances in 2D, 3D and higher-order topological photonics. *Light Sci. Appl*. **9**, 130 (2020).

[24] D. a. Bisharat, R. Davis, Y. Zhou, P. Bandaru, and D. Sievenpiper. Photonic Topological Insulators: A Beginner's Introduction [Electromagnetic Perspectives]. *IEEE Antennas and Propagation Magazine*. **63**, 112 (2021).

[25] J. Chen and Z.-Y. Li. Topological photonic states in gyromagnetic photonic crystals: Physics, properties, and applications. *Chin. Phys. B*. **31**, 114207 (2022).

[26] G. J. Tang, X. T. He, F. L. Shi, J. W. Liu, X. D. Chen, and J. W. Dong. Topological Photonic Crystals: Physics, Designs, and Applications. *Laser & Photonics Reviews*. **16** (2022).

[27] N. Han, X. Xi, Y. Meng, H. Chen, Z. Gao, and Y. Yang. Topological photonics in three and higher dimensions. *APL Photonics*. **9** (2024).

[28] H. Wang, S. K. Gupta, B. Xie, and M. Lu. Topological photonic crystals: a review. *Front Optoelectron*. **13**, 50 (2020).

[29] D. Liu, P. Peng, X. Lu, A. Shi, Y. Peng, Y. Wei, and J. Liu. Multiple topological states within a common bandgap of two non-trivial photonic crystals. *Opt. Lett*. **49**, 2393 (2024).

[30] B. Yan *et al.* Multifrequency and Multimode Topological Waveguides in a Stampfli-Triangle Photonic Crystal with Large Valley Chern Numbers. *Laser & Photonics Reviews*. **18**, 2300686 (2024).

[31] Z. Wang, Y. D. Chong, J. D. Joannopoulos, and M. Soljačić. Reflection-free one-way edge modes in a

gyromagnetic photonic crystal. *Phys. Rev. Lett.* **100**, 013905 (2008).
[32] Z. Wang, Y. Chong, J. D. Joannopoulos, and M. Soljačić. Observation of unidirectional backscattering-immune topological electromagnetic states. *Nature*. **461**, 772 (2009).
[33] Z. Yu, G. Veronis, Z. Wang, and S. Fan. One-way electromagnetic waveguide formed at the interface between a plasmonic metal under a static magnetic field and a photonic crystal. *Phys. Rev. Lett.* **100**, 802 (2008).
[34] X. Ao, Z. Lin, and C. T. Chan. One-way edge mode in a magneto-optical honeycomb photonic crystal. *Phys. Rev. B*. **80**, 033105 (2009).
[35] T. Ochiai and M. Onoda. Photonic analog of graphene model and its extension: Dirac cone, symmetry, and edge states. *Phys. Rev. B*. **80**, 155103 (2009).
[36] Y. Poo, R.-x. Wu, Z. Lin, Y. Yang, and C. T. Chan. Experimental realization of self-guiding unidirectional electromagnetic edge states. *Phys. Rev. Lett.* **106**, 093903 (2011).
[37] A. B. Khanikaev, S. H. Mousavi, W.-K. Tse, M. Kargarian, A. H. MacDonald, and G. Shvets. Photonic topological insulators. *Nat. Mater.* **12**, 233 (2012).
[38] L.-H. Wu and X. Hu. Scheme for Achieving a Topological Photonic Crystal by Using Dielectric Material. *Phys. Rev. Lett.* **114**, 223901 (2015).
[39] G.-G. Liu *et al*. Topological Chern vectors in three-dimensional photonic crystals. *Nature*. **609**, 925 (2022).
[40] Y. C. Zhou, Z. Q. Sun, H. S. Lai, X. C. Sun, C. He, and Y. F. Chen. Observation of Photonic Chern Metal With Bi-Chiral Edge Propagation. *Laser & Photonics Reviews*. 2400826 (2024).
[41] H. Li, R. Ge, Y. Peng, B. Yan, J. Xie, J. Liu, and S. Wen. High-order nonreciprocal add-drop filter. *Science China Physics, Mechanics & Astronomy*. **64**, 124211 (2021).
[42] Y. Yang and Z. H. Hang. Topological whispering gallery modes in two-dimensional photonic crystal cavities. *Opt. Express*. **26**, 21235 (2018).
[43] X.-C. Sun and X. Hu. Topological ring-cavity laser formed by honeycomb photonic crystals. *Phys. Rev. B*. **103**, 245305 (2021).
[44] Y. Chen, N. Gao, G. Zhu, and Y. Fang. Chiral topological whispering gallery modes formed by gyromagnetic photonic crystals. *Phys. Rev. B*. **108**, 195423 (2023).
[45] F. D. M. Haldane. Model for a Quantum Hall Effect without Landau Levels: Condensed-Matter Realization of the "Parity Anomaly". *Phys. Rev. Lett.* **61**, 2015 (1988).
[46] A. P. Schnyder, S. Ryu, A. Furusaki, and A. W. W. Ludwig. Classification of topological insulators and superconductors in three spatial dimensions. *Phys. Rev. B*. **78**, 195125 (2008).
[47] C. Fang, M. J. Gilbert, and B. A. Bernevig. Bulk topological invariants in noninteracting point group symmetric insulators. *Phys. Rev. B*. **86**, 115112 (2012).
[48] J. Zhu, Y. Liu, L. Jia, B. Zhang, Y. Yang, and D. Tang. High-frequency magnetodielectric response in yttrium iron garnet at room temperature. *J. Appl. Phys.* **123** (2018).
[49] J. Chen and Z.-Y. Li. Prediction and Observation of Robust One-Way Bulk States in a Gyromagnetic Photonic Crystal. *Phys. Rev. Lett.* **128**, 257401 (2022).
[50] Y.-C. Zhou, H.-S. Lai, J.-L. Xie, X.-C. Sun, C. He, and Y.-F. Chen. Magnetic corner states in a two-dimensional gyromagnetic photonic crystal. *Phys. Rev. B*. **107**, 014105 (2023).
[51] J. Ward and O. Benson. WGM microresonators: sensing, lasing and fundamental optics with microspheres. *Laser & Photonics Reviews*. **5**, 553 (2011).
[52] B. Bahari, A. Ndao, F. Vallini, A. El Amili, Y. Fainman, and B. Kanté. Nonreciprocal lasing in topological cavities of arbitrary geometries. *Science*. **358**, 636 (2017).
[53] B. Bahari, L. Hsu, S. H. Pan, D. Preece, A. Ndao, A. El Amili, Y. Fainman, and B. Kanté. Photonic quantum Hall effect and multiplexed light sources of large orbital angular momenta. *Nat. Phys.* **17**, 700 (2021).
[54] T.-F. Li, Y.-L. Li, Z.-Y. Zhang, Q.-H. Yang, F. Fan, Q.-Y. Wen, and S.-J. Chang. Terahertz faraday rotation of magneto-optical films enhanced by helical metasurface. *Appl. Phys. Lett.* **116**, 251102 (2020).
[55] Q. Xue, Y.-J. Zhang, D. Zhao, Q.-H. Yang, H.-W. Zhang, F. Fan, and Q.-Y. Wen. Enhanced terahertz magneto-optical performance in substrate-free ultra-thick TbErBi:RIG crystal films. *Appl. Phys. Lett.* **123**,

141903 (2023).

[56] F. Dirnberger *et al.* Magneto-optics in a van der Waals magnet tuned by self-hybridized polaritons. *Nature*. **620**, 533 (2023).

[57] Y.-P. Ruan *et al.* Observation of loss-enhanced magneto-optical effect. *Nat. Photonics*. (2024).

# Figures and captions

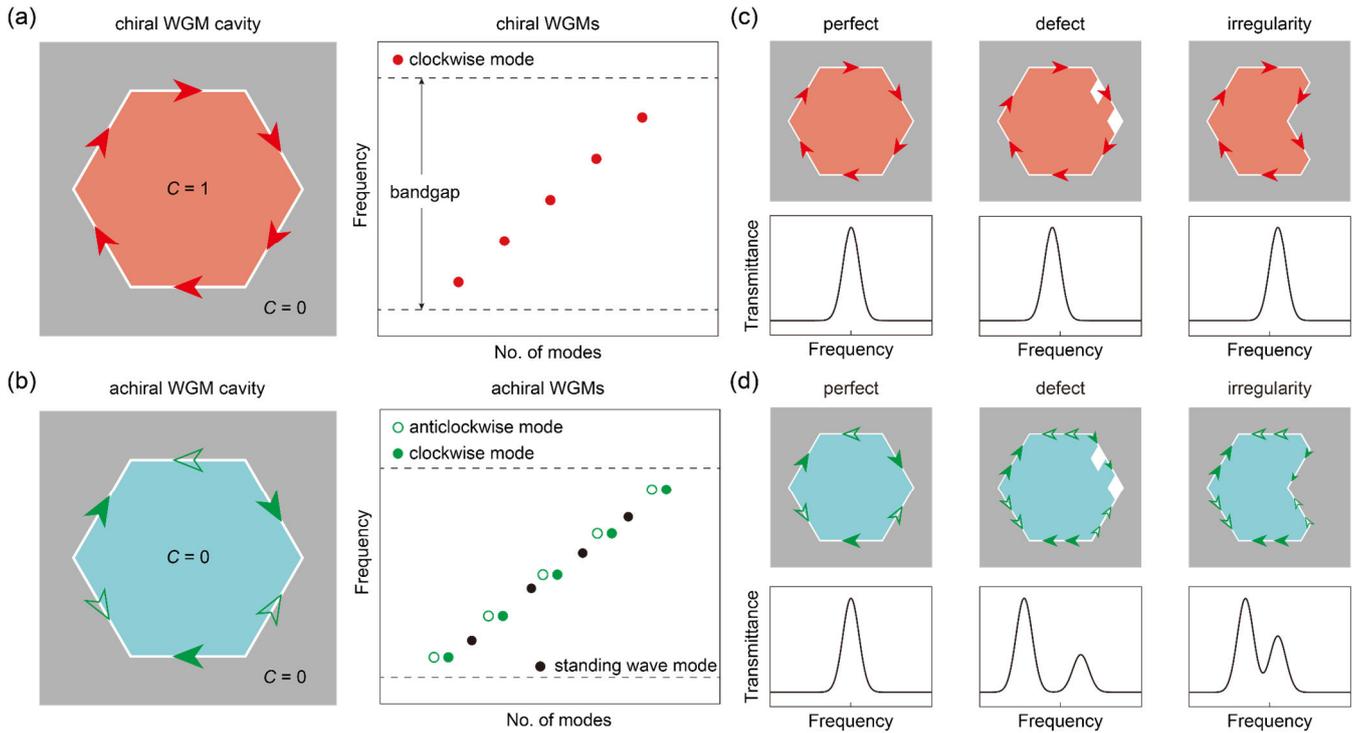

**FIG. 1. Chiral whispering gallery mode (WGM) cavity and achiral WGM cavity.** (a, b) Schematics of chiral and achiral WGM cavities. The chiral WGM cavity is constructed by the photonic Chern insulator with a nonzero Chern number, supporting chiral WGMs within the bandgap. In contrast, the achiral WGM cavity is constructed by the photonic ordinary insulator, supporting both clockwise and anticlockwise WGMs. (c, d) The effects of different defects and irregularity on both cavities. In the chiral WGM cavity, defects or irregularities only shift the cavity mode frequency, maintaining a single resonant peak. In contrast, in the achiral WGM cavity, similar defects or irregularities not only induce mode shift but also result in mode splitting.

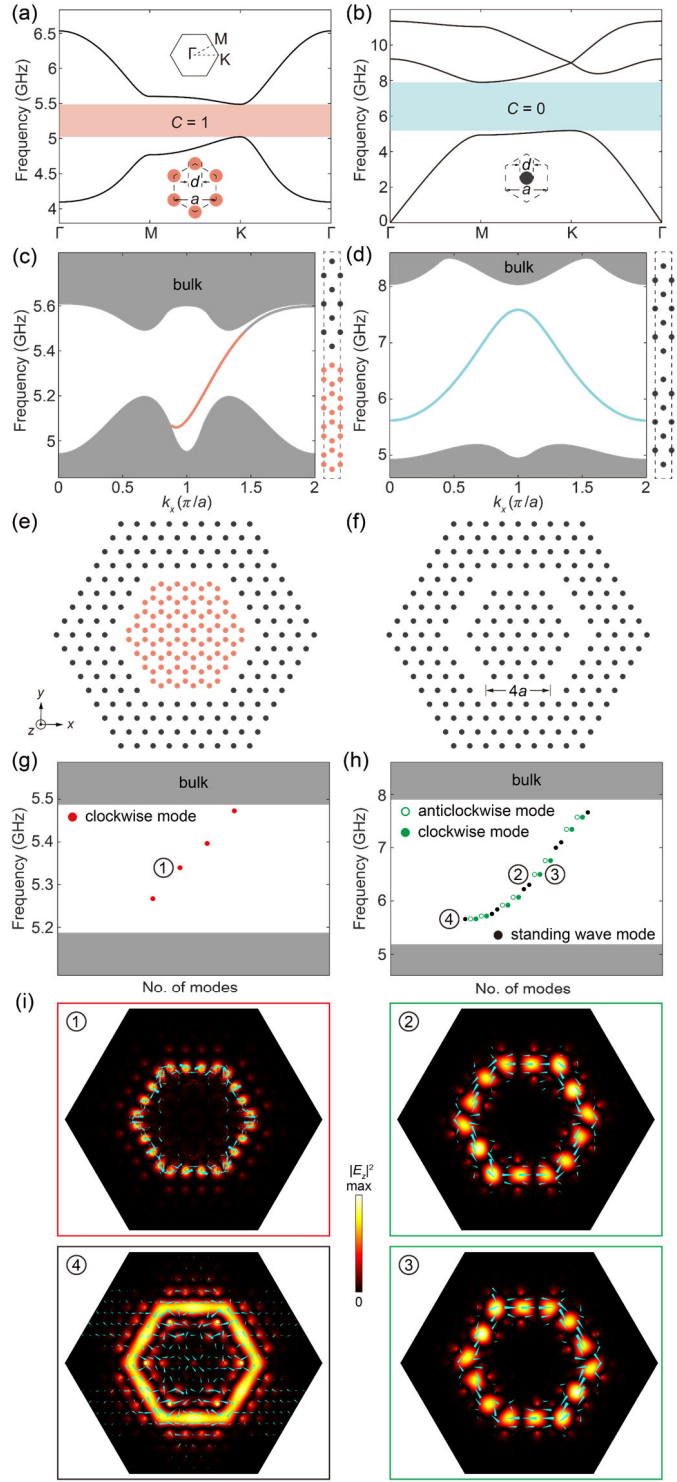

**FIG. 2. Design of chiral and achiral WGM cavities.** (a, b) Photonic bulk bands of (a) the photonic Chern insulator which is constructed by a honeycomb lattice of YIG rods with external magnetic fields [represented by pink circles] and (b) the photonic ordinary insulator which is constructed by a triangular lattice of YIG rods without the external magnetic fields [represented by black circles]. Both PCs have a lattice constant of $a = 20$ mm. (c, d) Edge dispersions of (c) chiral edge modes of the photonic Chern insulator and (d) achiral edge modes of the photonic ordinary insulator. The right insets show the morphology of corresponding edges. (e, f) Schematics of chiral and achiral WGM cavities. (g, h) The simulated WGMs of both cavities. Within the bandgap, only chiral WGMs are found in the chiral WGM cavity, while both clockwise and anticlockwise WGMs are observed in the achiral WGM cavity. (i) $|E_z|^2$ fields and Poynting vectors of representative WGMs of both cavities.

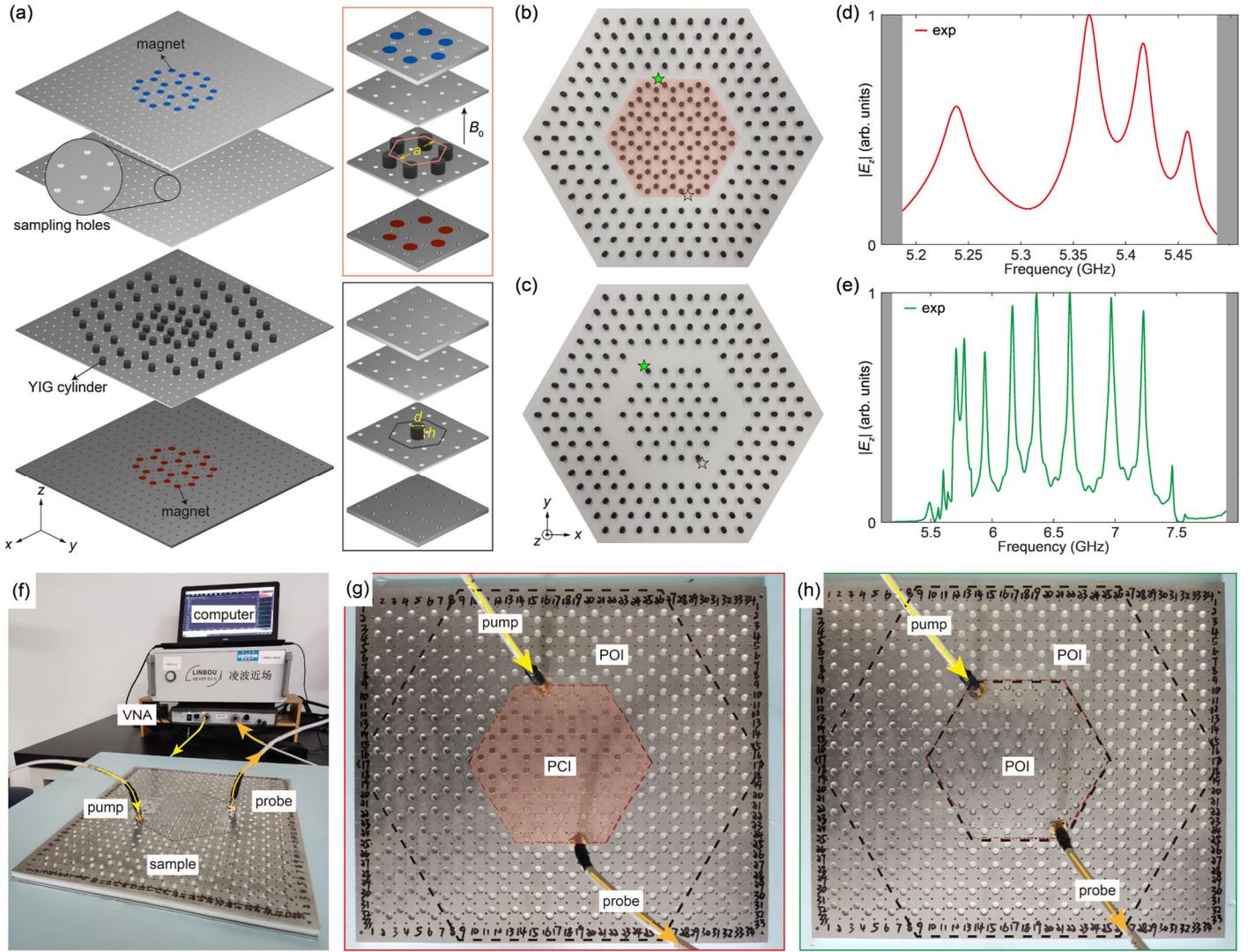

**FIG. 3. Realization and measured |$E_z$| spectra of chiral and achiral WGM cavities.** (a) Schematic of the experimental sample. The top and bottom magnets (marked in blue and red circles) apply external magnetic fields on the YIG cylinders. Periodic YIG cylinders are sandwiched between two metallic plates. Sampling holes are drilled in metallic plates to insert the source and probe antenna. The insets in pink and black boxes illustrate the experimental setups for YIG rods with and without the external magnetic fields, respectively. (b, c) Top-view photos of the fabricated chiral and achiral WGM cavities. The top metallic plates are removed for better visualization. The filled stars indicate the positions of source antenna, while the unfilled stars indicate the positions of probe antenna used to measure the |$E_z$| spectra. (d, e) Measured |$E_z$| spectra in (d) the chiral WGM cavity and (e) the achiral WGM cavity. (f) Experimental setup for pump-probe measurement. VNA: Vector Network Analyzer. (g, h) The positions of pump and probe antennas for chiral (g) and achiral (h) WGM cavities, respectively. The hexagons indicate the boundaries of POI and PCI used to construct the cavity.

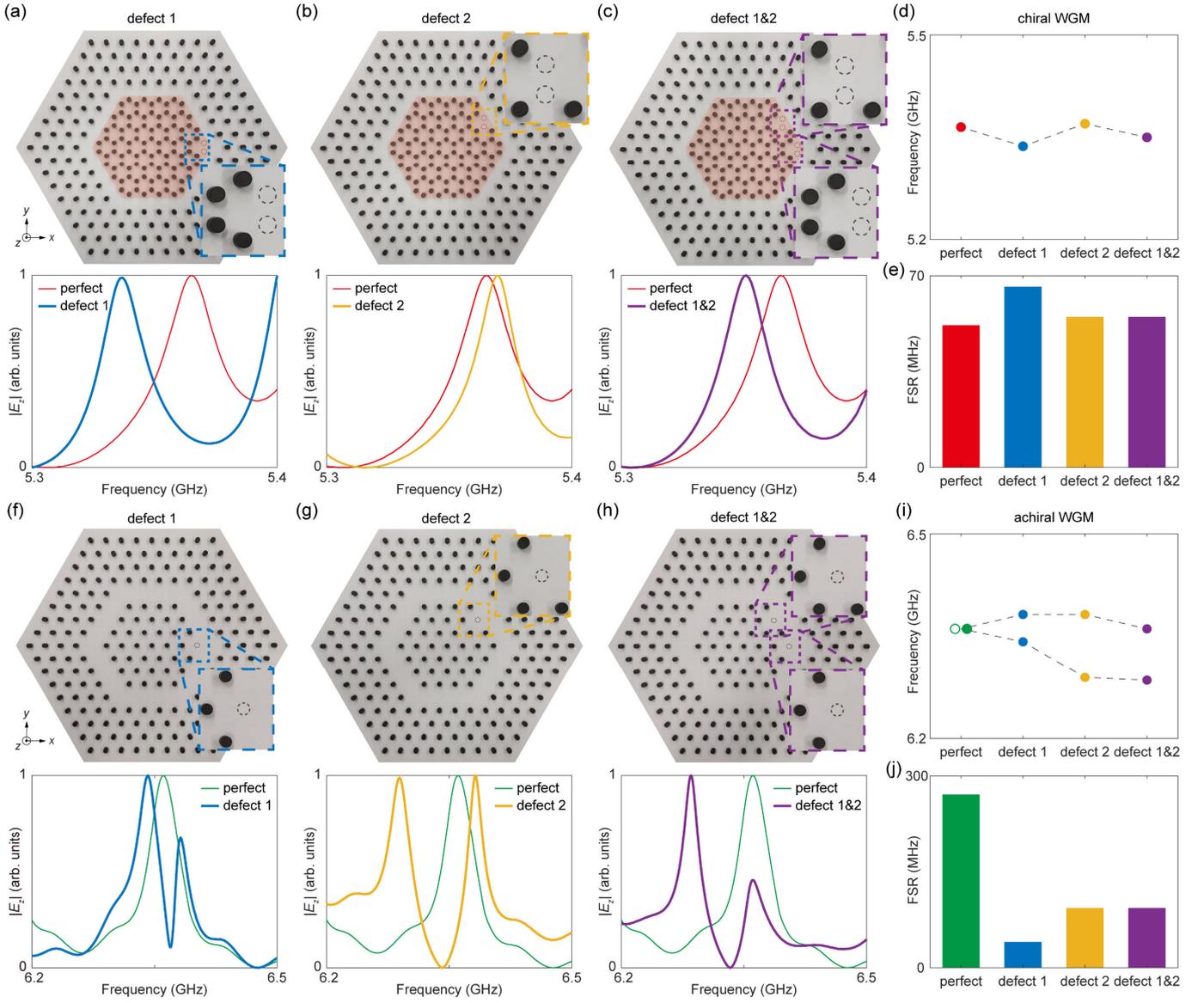

**FIG. 4. Response of the chiral and achiral WGM cavities to different defects.** (a-c) Chiral WGM cavities with three different defects. The first and the second defects is constructed by removing two YIG cylinders along the boundary but at different locations. The third defect is a combination of the first and second defects. Zoom-in pictures of defects are shown as insets in which dotted circles outline the removed rods. The measured $|E_z|$ spectra show one single peak but at different resonant frequency. (d) The resonant frequency of chiral WGM of four different cavities. (e) Free spectral range (FSR) of the chiral WGM. (f-j) Achiral WGM cavities with three different defects, the corresponding $|E_z|$ spectra, resonant frequency of WGM and its FSR.

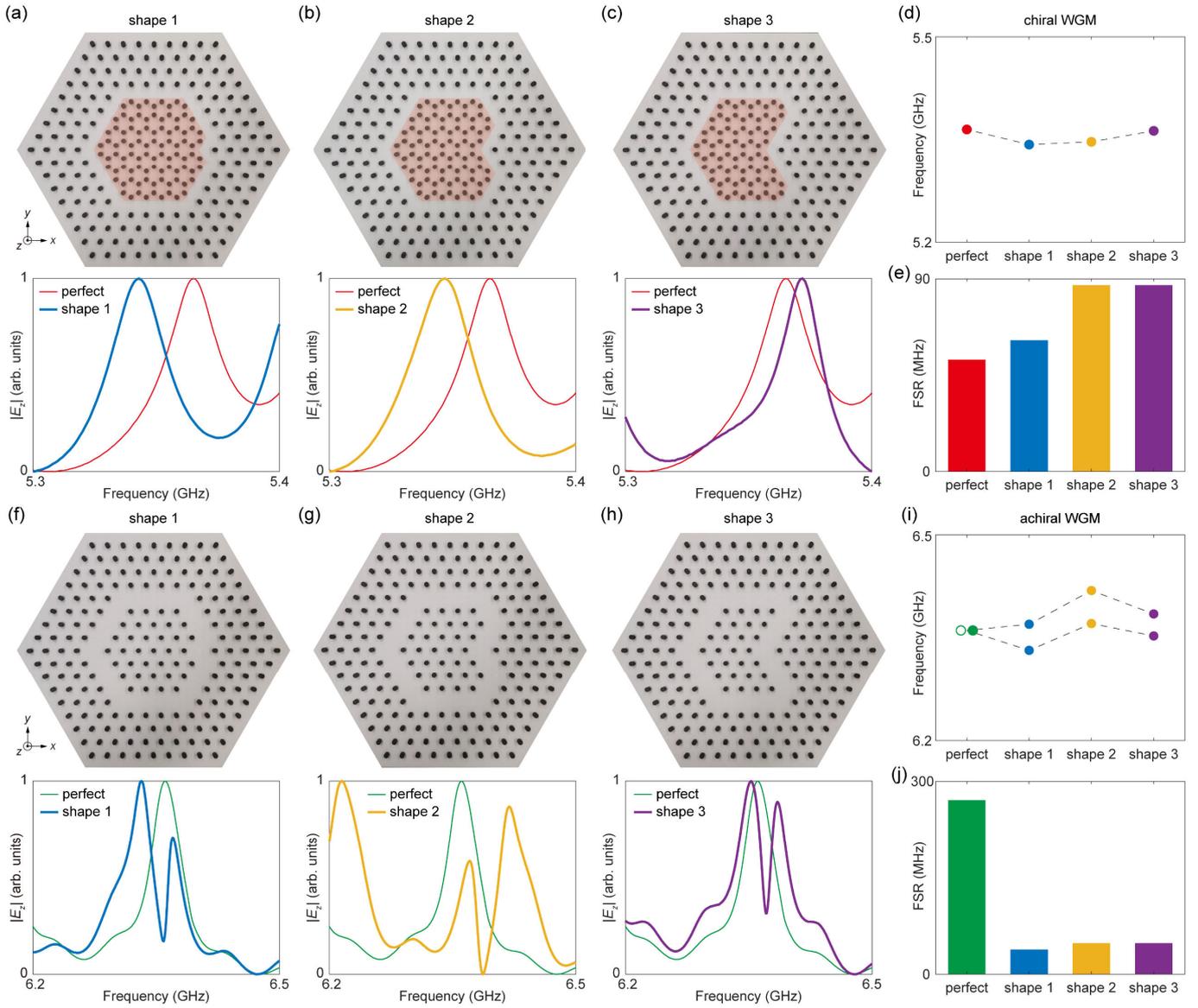

**FIG. 5. Response of the chiral WGM and achiral WGM cavities to different irregularities.** (a-c) Chiral WGM cavities of three different shapes. These shapes indicate a gradual increase in the irregularity of the cavity. One single peak is observed at different resonant frequencies within each measured $|E_z|$ spectrum. (d) The resonant frequency of chiral WGM of four different cavities. (e) FSR of the chiral WGM. (f-j) Achiral WGM cavities of three different shapes, the corresponding $|E_z|$ spectra, resonant frequency of WGM and its FSR.